\documentclass[structabstract]{aa}  
\usepackage{graphicx}
\usepackage{txfonts}
\usepackage{textcomp}

\catcode`\@=11
\def\gsim{\ifmmode{\mathrel{\mathpalette\@versim>}}
    \else{$\mathrel{\mathpalette\@versim>}$}\fi}
\def\lsim{\ifmmode{\mathrel{\mathpalette\@versim<}}
    \else{$\mathrel{\mathpalette\@versim<}$}\fi}
\def\@versim#1#2{\lower 2.9truept \vbox{\baselineskip 0pt \lineskip 
    0.5truept \ialign{$\m@th#1\hfil##\hfil$\crcr#2\crcr\sim\crcr}}}
\catcode`\@=12

\begin{document}
   \title{3D kinematics through  the X-shaped Milky Way bulge
   \thanks{Based on observations taken with ESO telescopes at the La Silla Paranal 
   Observatory under programme IDs 163.O-0741(A), 085.D-0143(A) and 385.B-0735(B); 
   and with observations taken with the Magellan telescope at the Las Campanas 
   Observatory.}}

\author{
 S. V\'{a}squez\inst{1,2,3} 
 \and
 M. Zoccali\inst{1,3}
 \and
 V. Hill\inst{4}
 \and 
 A. Renzini\inst{5,6}
 \and
 O. A. Gonz\'{a}lez\inst{2,3}
 \and           
 E. Gardner\inst{7}
 \and
 Victor P. Debattista\inst{8}
 \and
 A. C. Robin\inst{7}
 \and 
 M. Rejkuba\inst{9}
 \and
 M. Baffico\inst{1}
 \and
 M. Monelli\inst{10,11}   
 \and 
 V. Motta\inst{12}         
 \and 
 D. Minniti\inst{1,3,13,14}
}

\institute{
Departamento de Astronom\'{i}a y Astrof\'{i}sica, Pontificia Universidad Cat\'olica 
de Chile, Av. Vicu\~na Mackenna 4860, Santiago, Chile.
\email{svasquez@astro.puc.cl; mzoccali@astro.puc.cl}
\and
European Southern Observatory, Alonso de Cordova 3107, Santiago, Chile
\and
The Milky Way Millennium Nucleus, Av. Vicu\~{n}a Mackenna 4860, 782-0436 Macul, 
Santiago, Chile.
\and
Laboratoire Lagrange UMR7293, Universit\'e de Nice Sophia--Antipolis, CNRS, 
Observatoire de la C\^{o}te d'Azur, BP4229, F-06304 Nice, France
\and 
INAF - Osservatorio Astronomico di Padova, vicolo dell'Osservatorio 5, 35122, Padova, 
Italy.
\and
GEPI, Observatoire de Paris, CNRS UMR 8111, Universit\'{e} Paris Diderot, F-92125, Meudon,
Cedex, France.
\and
Institut Utinam, CNRS UMR 6213, OSU THETA, Universit\'{e} de Franche--Comt\'{e}, 41bis avenue de l'Observatoire, 25000 Besan\c{c}on, France
\and
Jeremiah Horrocks Insitute, University of Central Lancashire, Preston, PR1 2HE, UK.
\and
European Southern Observatory, Karl-Schwarzschild-Strasse 2, 85748, Garching, Germany.
\and
Instituto de Astrof\'{i}sica de Canarias, Calle V\'{i}a L\'actea s/n, E-38205 La Laguna, Tenerife, Spain.
\and
Departmento de Astrof\'{i}sica, Universidad de La Laguna, E38200 La Laguna, Tenerife, Spain.
\and
Departamento de F\'{i}sica y Astronom\'{i}a, Universidad de Valparaiso, Avenida Gran Bretana 1111, Valparaiso, Chile.
\and
Vatican Observatory, V00120 Vatican City State, Italy.
\and 
Departamento de Ciencias F\'{i}sicas, Universidad Andres Bello, Santiago, Chile
}
   \date{Received; accepted}

 
\abstract
{It has recently been discovered that the Galactic bulge is X-shaped,
  with the two southern $arms$ of the X both crossing the lines of
  sight at $l=0$ and $|b|>4$, hence producing a double red clump in
  the bulge color magnitude diagram. Dynamical models predict the
  formation of X-shaped bulges, as extreme cases of boxy-peanut
  bulges. However, since X-shaped bulges were known to be present only
  in external galaxies,
  models have never been compared to kinematical data for individual
  stars.}
{We study the orbital motion of Galactic bulge stars, in the two $arms$
(overdensities) of the X in the southern hemisphere. The goal is to provide 
observational constraints to bulge formation models that predict the
formation of X-shapes through bar dynamical instabilities.}
{Radial velocities have been obtained for a sample of 454 bulge giants, 
roughly equally
distributed between the bright and faint red clump, in a field at 
($l,b$)=($0,-6$).
Proper motions were derived for all red clump stars in the same 
field by combining images from two epochs  obtained 11 years apart, 
with WFI@2.2m at La Silla. The observed field contains the Globular 
Cluster NGC 6558, whose member stars were used to assess the accuracy 
of the proper motion measurement. At the same time, as a by product, we 
provide the first proper motion measurement of NGC 6558.
The proper motions for the spectroscopic sub-sample are analyzed 
taking into account the radial velocities and metallicities 
measured from near-IR Calcium triplet lines, for a subsample of 352 stars.}
{The radial velocity distribution of stars in the bright red clump, tracing the
closer overdensity of bulge stars, shows an excess of stars moving towards 
the Sun. Similarly, an excess of stars receding from the Sun is seen in the far
overdensity, traced by faint red clump stars. This is explained with 
the presence of stars on  elongated orbits, most likely streaming 
along the arms of the X-shaped bulge. Proper motions for these stars 
are consistent with 
qualitative predictions of dynamical models of peanut--shaped bulges. 
Surprisingly, stars on elongated orbits have preferentially metal 
poor (subsolar) metallicities, while the metal rich ones, in 
both overdensities, are preferentially
found in more axysimmetric orbits.

The observed proper motion of NGC 6558 has been measured as  
$(\mu_{\rm  l}\cos(b),\mu_{\rm b})=(0.30 \pm 0.14, -0.43\pm
0.13)$, with  a dispersion of  $(\sigma_l \cos(b),\sigma_b)=(1.8,1.7)$
mas/yr. Being the first PM measurement for this cluster.
}
{}

   \keywords{ Galaxy: bulge -- Galaxy: kinematics and dynamics -- Galaxy: structure -- Galaxies: kinematics and dynamics -- globular clusters: individual: NGC 6558
               }

   \maketitle
%

\section{Introduction}

With  $\sim  10^{10} M_\odot$  in stars  (Kent 1992)  the Galactic
bulge is, after the disk, the second most massive stellar component of
the Milky  Way. It is the only galactic bulge in which its  individual stars
can be resolved down to the  bottom of the main sequence, allowing the
construction  of  deep  color-magnitude  diagrams  (CMDs),  astrometric
proper motions measurements, radial velocities and detailed
chemical   composition  studies  from   medium  and   high  resolution
spectroscopy. Thus, the Milky Way bulge offers a unique opportunity to
map the stellar content of  a bulge, as summarized by age, metallicity
and  kinematical   multivariate  distributions.  All   together,  this
observational  evidence should  help us to reconstruct  the formation
history of  the bulge, hence identify the  basic physical processes
that have led to its present dynamical structure and stellar content.

At  least  three  distinct  processes  have  been  envisaged  for  the
formation  of  galactic bulges.   An  early  formation  by merging  of
gas-rich  smaller galaxies,  a  process traditionally  invoked for  the
formation of  elliptical galaxies (Toomre 1977); a  late formation via
the secular growth of a bar  instability in a pure stellar disk (e.g.,
Combes  \& Sanders  1981; Saha  et  al. 2010;  Shen et  al. 2010,  and
references therein); and, more  recently, an early formation via clump
instability, migration and central coalescence in a very gas-rich disk
(e.g.  Immeli et  al. 2004;F\"{o}rster Schreiber et al. 2006 ; 
Carollo et  al.  2007;  Elmegreen  et al. 2008; Bournaud et  al. 2009; 
Genzel et al. 2011). Moreover, secular processes  driven by a
bar may  dynamically change a  pre-existing non-rotating bulge  into a
boxy bulge with high cylindrical rotation (Saha, Martinez-Valpuesta \&
Gerhard  2012). The  predominantly old  stellar content  of  the bulge
($\gsim 10$ Gyr, Ortolani et  al. 1995; Zoccali et al.  2003; Clarkson
et al. 2011) demands an early formation, hence favoring processes that
are supposed  to act at an early  time. On the other  hand, the proven
bar shape of the Galactic bulge (Stanek et al. 1994; Dwek et al. 1995;
Rattenbury et al.  2007;  and references therein) clearly demands that
some sort of  disk and bar instability had taken  place. Thus, it is
quite  natural  to  expect  that  a  variety  of  processes  may  have
contributed to the build up of the bulge.

Although there has been general agreement that the bulge is bar-shaped,
some of  its structural parameters  are still under  discussion. Its axial ratios appear to be close to 1:0.35:0.25 but its inclination
angle w.r.t. the line of sight  has been reported in the range between
$\sim 15$  and $45$ degrees, depending  on the method to  trace it
(e.g.,  Binney, Gerhard  \&  Spergel 1997;  Dehnen  2000; Bissantz  \&
Gerhard  2002;  Benjamin et  al.   2005;  Babusiaux  \& Gilmore  2005;
Rattenbury  et al.   2007; Robin et al. 2012; and  references therein).   
Star  counts at
longitudes $|l|>7$ provided hints  for the presence of a second,
longer and  thinner bar (e.g., Hammersley et  al. 2000; Cabrera-Lavers
2007, 2008;  L\'opez-Corredoira et al. 2007; Churchwell  et al. 2009),
which however has been interpreted as more likely being a component of
the bar itself (e.g., Martinez-Valpuesta \& Gerhard 2011; Athanassoula
2012).

More recently,  McWilliam \& Zoccali  (2010) and Nataf et  al.  (2010)
used star counts  of red clump (RC) stars from  the 2MASS (Skruskie et
al.    2006)  and   OGLE-III   (Szymanski  et   al.   2011)   catalogs,
respectively, to  show that, along the $l=0$  direction, for latitudes
exceeding  $|b|=5$, the RC  splits in  two components,  indicating the
presence  of two peaks  in stellar  density along  the line  of sight.
McWilliam \&  Zoccali (2010) interpreted  this feature as evidence
for  the Galactic  bar being  X-shaped, which  was later  confirmed by
Saito  et al.  (2011)  in a  more quantitative  analysis of  2MASS RC
giants across a  larger bulge area.  In this context,  the main bar is
just seen as the inner part of the X-shaped bulge.

X-shaped bulges can be qualitatively produced by some dynamical models
as   peculiar  boxy/peanut (B/P)   structures   (e.g.,  Athanassoula   2005;
Martinez-Valpuesta, Shlosman \& Clayton  2006, Debattista et al. 2006)
and   X-shaped   isophotes   have   been   observed  in   a   few   S0
galaxies.   However,  such   models  have   not  been   compared  with
observations of  velocities for samples of individual  stars, nor have
been fine--tuned to reproduce  a bulge with the characteristics of
the Milky  Way bulge, at least in part due to lack of observational 
constraints. We  are now beginning  to collect kinematical data  
for a large number of  stars in the Galactic
bulge, thus setting constraints on the proposed formation mechanisms.

Proper motions (PMs)  can be obtained for large  samples of stars with
relatively short exposure times, but  a long time baseline and precise
astrometry are required in order to reach the needed accuracy of a few
mas/yr, corresponding to  the bulge proper motion dispersion. A
pioneer study based on photographic  plates taken 33 years apart has 
yielded proper motions  for  $\sim  400$  stars in  Baade's  Window
(Spaenhauer et al. 1992), later followed by Rich \& Terndrup (1997) in
the same field.  Zoccali et al.  (2001) used the WFPC2 camera on board
the Hubble Space  Telescope to measure the proper  motion of the bulge
globular cluster  NGC~6553 as well  as the bulge dispersion  of proper
motions in the  same field, and so did Feltzing  \& Johnson (2002) for
the bulge  globular cluster NGC~6528.  Vieira et  al.  (2007) measured
proper motions in  the Plaut Window at $(l,b)=(0,-8)$  whereas a large
proper motion catalog based on  OGLE-II data, extending over several
fields  mostly at  $b=-4$,  has  been obtained  by  Sumi et  al.
(2004). This catalog  was used by Rattenbury et  al. (2007) to compare
the   PM  distribution  of   RC  stars   with    bulge  formation
models.  Kuijken \& Rich  (2002) used  the WFPC2  camera on  board the
Hubble  Space Telescope  to  demonstrate  that the  bulge  CMD can  be
decontaminated  from the foreground  disk stars  based on  the different
proper motion  distributions of bulge and disk  stars.  Following this
approach, Clarkson et al.  (2008, 2011) used two epoch ACS/HST data to
derive  a  bona  fide  pure-bulge   CMD  for  a  field  at  $(l,b)\sim
(1,-2.51)$,  clean  from  disk  contamination, demonstrating  that the
bulge consists  of  a predominantly  old  population  of  $\gsim 10$  Gyr,  
and confirming a  similar conclusion by Ortolani et al. (1995) and 
Zoccali  et al. (2003)  who used a more crude or a statistical foreground decontamination, respectively.

Important  constraints  on the  bulge  formation  mechanism come  from
radial velocity  measurements (Rich 1988, 1990; Terndrup  et al. 1995;
Minniti  et al.   1996; Sadler  et al.   1996) especially  when proper
motions are available on the same  field.  In fact, Zhao et al. (1996)
combined  the proper  motions by  Spaenhauer et  al.  (1992)  with the
radial velocities from most of the above mentioned samples deriving 3D
space  velocities  for  a  sample  of  62  K  giants,  that  showed  a
significant vertex  deviation.  This result, later  confirmed by Soto,
Rich \& Kuijken  (2007), indicates a bar-like  structure for the
Galactic  bulge.   With  a   complex,  simultaneous  analysis  of  the
metallicity  distribution function  and kinematics  of  Baade's Window
giants, Babusiaux et al.  (2010)  and Hill et al.  (2011) demonstrated
that the  vertex deviation is mainly  produced by the  most metal rich
stars  which have  disk-like $\alpha$-element  abundance  (Gonzalez et
al.  2011),  while  the  metal  poor, $\alpha$-enhanced  stars  have  a
kinematics  more typical  of a  classical spheroid.  The  BRAVA survey
(Howard  et al.  2008,  2009;  Kunder et  al.   2012) obtained  radial
velocities  for a total  of $\sim  10,000$ bulge  M giants  in several
fields  spanning  a  wide  range   in  longitude  ($-10  <  l  <
+10$)  and latitudes  ($b=-4,  -6$ and  $-8$),
finding evidence  for a cylindrical rotation. By  comparing those data
with their  simple N-body  model, Shen et  al.  (2010) argued  for the
Milky Way being a pure disk galaxy, i.e., without the need of a merger
made bulge. It should be emphasized that the vast majority of 
early--type galaxies (over $\sim$ 86\%, cf. Emsellem et al. 
2011) are actually {\it fast rotators}, with just the 
most massive  elliptical galaxies 
being predominantly  {\it slow rotators}. Thus, the cylindrical rotation 
of the bulge does not necessarily imply a formation mechanism radically 
different from that of the majority of early-type galaxies.

A  different  approach of combining metallicity  with radial  velocity
measurements has been followed by  Rangwala et al.  (2009a,b) who used
Fabry-Perot imaging to  sample the Calcium II Triplet  (CaT) lines for
$\sim 3,400$  stars along the  bar ($l=0,\pm5$) at  $b=-4$. They
detected  the presence  of  bar stellar  streaming  motions along  the
spanned  longitudes.  Radial  velocities and  metallicities separately
for bright and faint red clump stars were first obtained by de Propris
et al.  (2011)  in a bulge field at  $(l,b)=(0,-8$), 
and more  recently by  Uttenthaler  et al.  (2012) in  a field  at
$(l,b)=(0,-10)$, and  by Ness et al  (2012) in three  fields along the
minor axis at $b<-5$.

In  the present  paper we  combine radial  velocity and  proper motion
measurements to derive the 3D motion  of 454 bulge stars in the bright
and faint red clump of a field at $(l,b)=(0,-6)$. Like those in the de
Propris et al. study, these stars trace the near and far {\it arms} of
the  X-shaped   bulge,  respectively.   The  paper   is  organized  as
follows. In Section  2 we describe the data  and reduction methods for
our  photometric, astrometric and  spectroscopic measures,  whereas in
section 3 and 4 the resulting radial velocities and proper motions are
presented. In section 5 are presented the space velocities. 
The metallicity  distributions of the stars in  the two red
clumps  are  presented in  Section~6,  and  conclusions  are drawn  in
Section~7.


\section{Observations and Data Reduction}

\begin{figure}
   \centering
   \includegraphics[width=9cm]{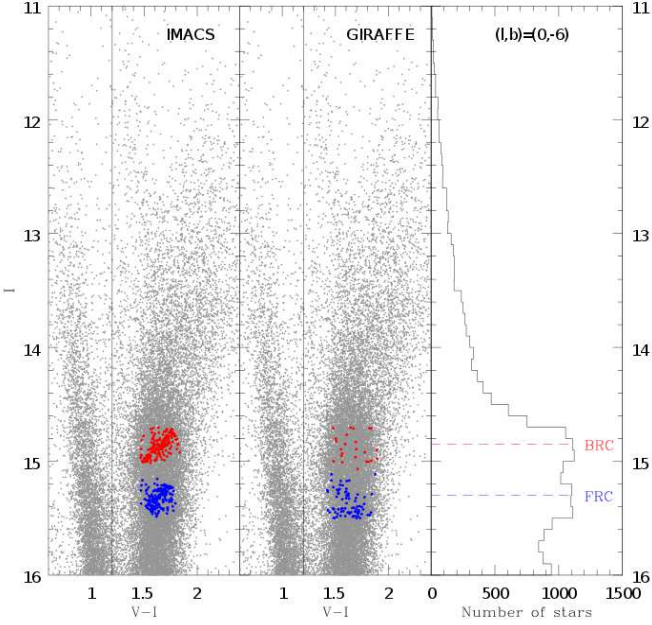}
      \caption{Left and middle panels: Optical color magnitude diagram 
      of the bulge field at $(l,b)=(0,-6)$ obtained with WFI@2.2m at 
      La Silla. The vertical line marks the color cut applied to 
      construct the histogram on the right ($V-I>1.2$), where the two 
      clumps are clearly visible. The Bright Red Clump (BRC -- red points) 
      and Faint Red Clump (FRC -- blue points) spectroscopic target selection 
      observed with IMACS on Magellan and with FLAMES-GIRAFFE on VLT 
      are overplotted on the CMD in the left and middle panels, respectively.
      }
      \label{cmd}
   \end{figure}

\subsection{WFI Photometry}

The  spectroscopic targets were  selected based  on the  optical $V,I$
photometry of a bulge  field centered at $(l,b)=(0,-6)$, obtained with
the  WFI camera  at  the 2.2m  telescope  at ESO  La  Silla, on  April
15$^{th}$, 1999 as part of the ESO imaging Survey ESO programme (EIS, 
ESO programme ID 163.O-0741(A)).  The resulting CMD has
been  already presented  in Zoccali  et al.   (2003) and  McWilliam \&
Zoccali (2010). Figure~\ref{cmd} shows  the bright portion of the CMD,
which clearly displays  double red clump, as illustrated  in the right
panel.  Given the  modest dependence  of  the red  clump magnitude  on
stellar  population  properties (age  and  metallicity), McWilliam  \&
Zoccali concluded that the  two clumps trace two stellar overdensities
at  different  distances along  the  line  of  sight, and  that  these
overdensities  correspond to  the near  and  far arm  of the  X-shaped
bulge.  Based  on this CMD the  red clump targets  were selected among
the  bright (red  dots)  and faint  red  clump stars  (blue dots)  for
spectroscopic  follow-up with  Magellan IMACS  and  VLT FLAMES-GIRAFFE
spectrographs.

In  order to  derive space  velocities for  our targets,  second epoch
images of the same field were  obtained on May 6$^{th}$, 2010 with the
same instrument and filters   as in  the first epoch
(ESO programme  ID 085.D-0143(A)).  Thus, images were  taken through the
$V$- and $I$-band filters,  with integration times of 7$\times$50s for 
both.  While the  seeing of the first epoch was
  excellent (0.6--0.7 arcsec, observed at airmass=1.06), the  second epoch had 
  average seeing 1.6
  arcsec, at airmass=1.19.   Photometry  was carried  out  with 
  the  DAOPHOTII/ALLFRAME
  packages  (Stetson  1988,  1994),  on individual  chips,  while  the
  photometric calibration was derived  by comparison of stars with the
  first epoch.

\subsection{WFI Astrometry}

\begin{figure}
   \centering
   \includegraphics[width=9cm]{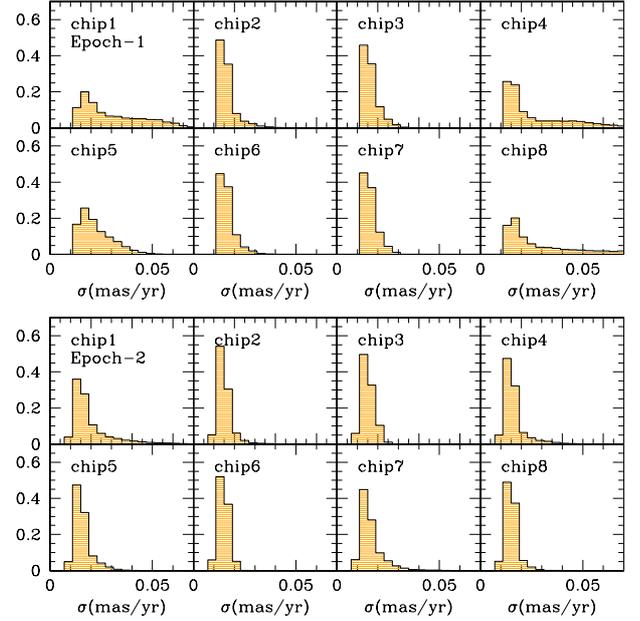}   
      \caption{The dispersion of the position ($X,Y$) of the stars in 
      every frame, for each epoch, is plotted across the WFI detector 
      in units of $mas/yr$. A higher spread 
      in the dispersion is observed in the edges of the CCD associated 
      with higher order distortion and variable  PSF. All of this effects are 
      mostly corrected in the procedure done in the astrometrization 
      (see text for details).
      }\label{sig_pix}
   \end{figure}

The $X,Y$ coordinates for each star were determined in each individual
exposure (hereafter  {\it frame}),  in each epoch. We used  7 frames in the first epoch (3 V and 4 I, see Zoccali et al. (2003) for details) and 14 frames in the second epoch (7 in V and I).   The set  of codes
developed  by  P.B.   Stetson (DAOMATCH and  DAOMASTER;  Stetson,  private
communication) were used  to transform the coordinates of  each star in
each frame  into the system  of the reference  frame of each  epoch. A
cubic transformation  was  allowed among different frames of each
epoch, in order to properly take into account distortions at the edges
of the chips. Different effects contribute to the distortions among different frames and between the two epochs. Among them, the non perfect alignment among the chips of the WFI mosaic shown in Fig. 2.7 of the WFI Handbook (version 2.2, March 2013) and the differential atmospheric refraction described in Filippenko (1982). Following Filippenko, we calculated the effect of atmospheric refraction, at the central wavelengths of the V and I filters, at the center of the detector and at one edge (15 arcmin away). The differential shift between the center and the edge varies by 0.002 arcsec between the two filters, at the airmass of epoch1 and by 0.0025 arcsec at the airmass of epoch2. This values would not be completely negligible in our case, because it corresponds to 0.01 pixels. However, this effect has been removed by the cubic coordinate transformation performed to each chip. An  atmospheric effect that can not be corrected by our procedure is the atmospheric dispersion. Due we are using broad band filters, the refracted light from a star will be dispersed along the parallactic angle producing a small spectra instead a point source image on the detector, where the maximum of the light distribution will depends on the color of the star. This fact will affect the determination of the spatial centroid, especially for stars with large color difference. In order to estimate this effect we compute the differential atmospheric refraction like above, for two stars with effective temperature of $T_{eff}=4500$ K and $T_{eff}=6000$ K (which correspond with a typical RC and MS star). The comparison shows that for both filters, the differential shift between the two stars are quite small, reaching $0.001$ mas/yr for V band and less than $0.0001$ mas/yr for I band. This scatter on the coordinates determination can not be corrected, but it is so small that will no affect our analysis.

The final catalogue  for each epoch was then obtained by
averaging the  positions of each star,  in all the frames  in which it
was detected.   Stars in the  magnitude range of interest  here $(I<16)$
were  detected in all the frames of the two epochs, with very few 
exceptions. An indication  of the
relative precision of our astrometry is given by the standard deviation, 
$\sigma$, of the
position  of a  given star  in different  frames.  This  is shown  in
Fig.~\ref{sig_pix},  where histograms  of $\sigma=(\sigma_{x}^{2}
  +\sigma_{y}^{2} )^{1/2}$, in  units of $mas/yr$ are plotted  for every WFI
chip,  in both  epochs. The  dispersion  is extremely  small for  all
chips,  typically within 0.05 mas/yr,  corresponding to  0.002
pixels. We can reach this small dispersion because the photometry  
has been done  with ALLFRAME,  a code
that {\it ``makes  simultaneous  use  of  the  geometric  and  photometric
information  from  all  the  frames  of a  given  field  to  derive  a
self-consistent set of positions  and magnitudes...''} (Stetson 1994).
In other  words, the  dispersion in  Fig.~\ref{sig_pix} is  very small
because    the   individual    measurements    are   not    completely
independent. Yet, it gives an estimate at least of the relative precision
in different parts of the mosaic. The figure also shows that  the dispersion 
at the edge of
the  mosaic (chip  \#1,\#4,\#5  and \#8)  is higher  than  in the  middle,
because of the  higher order distortion and variable  PSF. This effect
is larger in the first epoch, most likely due to the smaller number of 
frames and because the stellar profiles
here  are   slightly  undersampled,   due  to  the   excellent  seeing
conditions. 

The astrometrization of the final $X,Y$ catalogues was done by 
means of the IRAF
routines {\it ccxymatch, ccmap} and {\it cctran}. A 7-order polynomial
transformation  was adopted, and  the PPMXL  catalog (Roeser  et al.
\cite{roeser10})  was used  as  reference. The  two astrometrized  WFI
catalogs were then matched using the {\it topcat} catalog-handling
package  (Taylor 2005),  but  the residuals  showed  trends and  steps
mostly  corresponding to  the edges  of the  individual  chips.  These
trends  did not  disappear by  changing the  order of  the astrometric
solution, and they were even  larger if the UCAC3 catalogue (Zacharias
et  al. 2010)  was used  as reference  instead of  the PPMXL.   A more
complex  procedure was  then  adopted  to eliminate these  residual
trends.

First,  an astrometric  solution (7$^{th}$  order polynomial  for each
chip) was found in order  to convert ($X_1,Y_1$) to ($RA_1,DEC_1$) and
($X_2,Y_2$)  to ($RA_2,DEC_2$).  This  was done  using only  bulge RGB
stars\footnote{Only stars  with $V-I>1.4$ and $V<18.5$  were used, and
  the  disk red clump  sequence sticking  out from  the bulge  RGB, at
  $V-I=1.6$, $V=15$  upward and to  the blue, was also  excluded.} and
the PPMXL catalogue  as reference.  The two $RA,DEC$  catalogs were then
matched  to each other.   Residual trends  were present,  as discussed
above, but  this matched  catalogue was  used only as  a first  step, to
combine the chips in a single mosaic and have the star pairs in hand.

A  new transformation was  then derived  between pixel  coordinates of
star pairs in the two  epochs, ($X_1,Y_1$) and ($X_2,Y_2$), and it was
applied to the latter, to bring them to the pixel coordinate system of
the 1$^{st}$ epoch ($X_2^1,Y_2^1$).  For this transformation, in order
to avoid  adopting a high order  polynomial that could  {\it flare} at
the edges  of the  chips, we  preferred to divide  the field  in small
boxes 300$\times$300 pixel  wide (71$\times$71 arcsec), and impose
a first order transformation (a plane) in  each  subfield.   Finally,  both
($X_1,Y_1$) and  ($X_2^1,Y_2^1$) were transformed  to ($RA,DEC$) using
the ${\it  same}$ astrometric  solution found in  the first  step, for
epoch 1 only.  This last  step  ensured that  any  spurious  trend in  the
astrometric  solution was  applied to  the pixel  coordinates  of both
epochs, which was  effective in removing completely the  trends in the
residual differences  between the coordinates  of the two  epochs, thus
allowing us to minimize the proper motion error. The final $(RA,DEC)$
of both epochs were transformed into galactic coordinates $(l,b)$ by
means of the {\it topcat} package.

\subsection{IMACS spectra}
Spectra for 177  bright RC stars and 175 faint  RC stars were observed
with the multi slit mode of the IMACS spectrograph at the Las Campanas
Observatory  on July 10$^{th}$, 2010.  
A 1200 lines/mm grating  was used, with a blaze angle of
26.7 degrees. This setup produces  spectra centered at $\sim 8500$ \AA
\space (the precise value depending on the position of the star in the
field  of  view)  with  a  resolution $R\sim$5000.  The  spectra  have
S/N$\sim$40 in  the Calcium Triplet (CaT)  region, yielding velocities
accurate to a few km/s.

The spectra were reduced using the COSMOS pipeline, provided by The
Carnegie Observatories. This pipeline processes the multi slit spectra
from IMACS applying bias and flat field corrections, wavelength
calibration and sky subtraction to each 2d--spectrum.
The final extraction to 1d--spectra and velocity measurement was done
with IRAF \textit{apall} and \textit{fxcor} tasks. As a test for the
wavelength calibration made for COSMOS, the sky lines in each spectra
were cross--correlated with a sky lines template from UVES (Hanuschik
\cite{hanuschik03}). The residual shifts found ($\sim |10|$ km/s) were
applied to individual spectra in order to set them at the geocentric
rest frame.

The   radial    velocities   of   our   sample    were   measured   by
cross--correlation against  a synthetic spectral template for a typical 
RGB star with $T_{eff}=4750$ K and [Fe/H]$=-1.3$ covering the
region  from $8350\,\AA$  until $8950\,\AA$,  were the  CaT  lines are
located. Due  the IMACS  CCD mosaic configuration,  with 8 chips  in a
square array  separated by  gaps, in  a few spectra  one of  the three
lines fell  in the gap. In  such cases the measurement  was made using
only the available section in  the spectra and the respective range in
the template.  The typical  error obtained from the cross--correlation
was $\sim2.0$  km/s without significant outliers  ($\sigma \sim 0.8$
km/s).   Finally, heliocentric  corrections were  calculated  with the
IRAF task {\it rvcorrect} and applied to all radial velocities.

\subsection{FLAMES-GIRAFFE spectra}

Red clump stars in the same field at $(l,b)=(0,-6)$ were also observed
with the multi-fibre FLAMES-GIRAFFE spectrograph at  the VLT in Medusa 
mode within
the ESO programme ID 385.B-0735(B) in service mode. Spectra were taken
with  the LR08  setup,  centered  at 8817  $\AA$  yielding a  spectral
resolution of R=6500 and average S/N$\sim$50.  These observations were
not especially  fine tuned  to discriminate between  the two  RCs, and
therefore  the target stars  were not  evenly distributed  between the
two RCs. Of 130  Medusa fibres 24 were allocated to the  bright RC, and 78
to the faint RC stars.  The middle panel in Figure~\ref{cmd} shows the
FLAMES-GIRAFFE target selection in the CMD.

GIRAFFE  spectra were  extracted  and wavelength  calibrated with  the
instrument  pipeline available  from ESO,  and radial  velocities were
measured via  cross--correlation with IRAF  \textit{fxcor} task, using
the same  template set as  for IMACS spectra.  As expected  by the
difference  in  resolution as well as from the usage of fibre vs. slit spectrograph,  the  typical  error  for  GIRAFFE  radial
velocity measurements was smaller than for IMACS observations 
($V_{\rm Helio}$ error $\sim 1$ km/s with dispersion $\sim 0.3$ km/s).

There are four stars in common  between  IMACS and  GIRAFFE observed data 
set. These stars  have a mean difference on  measured radial velocity
of  $\Delta V_{\rm  Helio}\sim4\pm2$  km/s, and were used to measure 
the systematic difference in [Fe/H] for both samples (Sect. 6).


\section{Radial Velocities for bulge Red Clump stars}

\begin{figure}
   \centering
   \includegraphics[width=9cm]{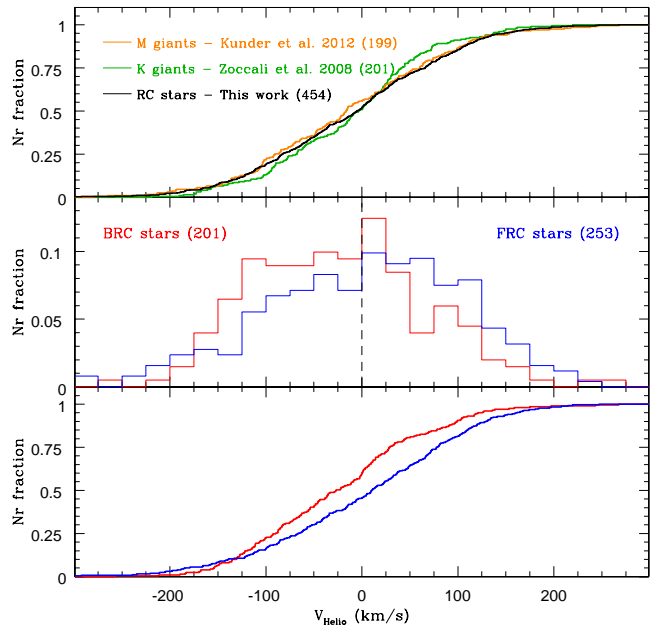}
      \caption{{\it Upper panel}: cumulative distributions of the 
        heliocentric radial velocities for bulge stars in the RC (black; 
        this work), K giants (green; Zoccali et al. 2008) and M giants (orange; 
        Kunder et al. 2012). All the stars are located in the same field at 
        galactic coordinates ($0,-6$). A Kolmogorov-Smirnov test shows
        that RC stars and M giants have radial velocity distributions that
        are similar, with a confidence of 71\%. On the contrary, K giants
        have a radial velocity distribution that is more peaked at
        $V_{\rm Helio}=0$ km/s with 89\% confidence of being {\it different} 
        from RC stars. See text for discussion.
        {Middle and lower panels}: Histograms and cumulative 
        distributions for heliocentric radial velocities of bulge stars in 
        the bright and faint RC, respectively.
        The two distributions differ from each other, with a confidence
        of 99.95\%, according to the Kolmogorov-Smirnov test.
      }\label{rv}
   \end{figure}

Figure  \ref{rv}  (top)   shows  the  cumulative  heliocentric  radial
velocity (RV) distribution  for the RC stars obtained  in this work,
compared with the sample of K giants stars analysed in
Zoccali et  al. (2008; see  also Babusiaux et  al. 2010) and  with the
sample of  M giants  observed within the  BRAVA survey (Kunder  et al.
2012).  The  three  samples  were   obtained  in  the  same  field  at
($l,b$)=($0,-6$), and the number of  stars in each sample is indicated
in the labels.  According to a Kolmogorov-Smirnov test, RC stars and M
giants have  the same RV distribution,  with a confidence  of 71\%. On
the other  hand, K giants, selected  in a box located  at $\sim 0.75$
mag above the RC, have  a RV distribution more peaked towards 
$V_{\rm Helio}=0$ km/s,
that  is  different  from  that  of  RC stars  with  a  confidence  of
89\%. This difference is also clear if we measure the RV dispersion of
the  two  samples:  $\sigma_{RC}=97\pm 4$~km/s  versus  $\sigma_{K}=83\pm
4$~km/s. The RV  dispersion of M giants from the  BRAVA sample, instead, is
$\sigma_{M}=101\pm  5$ km/s,  within $1\sigma$ of  that of  RC
stars.

One  possible   explanation  for   the  difference  between   the  RV
distribution of  RC and K giants,  in this field,  is a different
amount of contamination  by foreground disk stars, expected  to have a
RV distribution peaked  at $V_{\rm Helio}=0$ km/s.  Indeed, the disk  RC at different
distances and reddenings, along the  line of sight, is clearly visible
in the  CMD of Figure~\ref{cmd} as  a diagonal sequence  parallel to the
disk MS, but shifted $\sim$1  mag to the red. This sequence intersects
the bulge  RGB precisely  between $I=14$ and  $I=14.5$, which  are the
magnitude limits of the K giants target box (c.f., Figure~1 in Zoccali et al.
2008).  On  the contrary, M giants  in the BRAVA  sample were selected
such as  to avoid the disk  RC sequence, while bulge  RC stars largely
outnumber the disk foreground stars, in the same CMD box, because they
indeed {\it clump} at that magnitude.

The middle  and lower panels of Figure~\ref{rv}  show respectively the 
differential and cumulative heliocentric RV distribution of stars in the
bright and in the faint RC. Both the IMACS and GIRAFFE target stars are included in
the  histograms.  The Kolmogorov-Smirnov test  yields a confidence  of 99.95\%
that  the two  distributions are  different. In  fact, it  is visually
evident that there is an excess of stars with $V_{\rm Helio} \sim -80$
km/s in  the bright RC  sample, while a  similar excess is  visible at
$V_{\rm Helio} \sim  +80$ km/s in the faint RC  sample.  A dashed line
at $V_{\rm Helio}=0$  km/s has been drawn in  these panels to visually
emphasize the asymmetry of the two distributions.

Before  trying  to  understand   the  observed  asymmetry  in  the  RV
distribution of the two clumps, it is worth considering two sources of
contamination.   First of  all, we  know  that foreground  disk RC  at
distances $d>5$ kpc from the Sun would fall into the CMD selection boxes.
These  stars should have  RV close  to zero,  and a  narrower velocity
dispersion,  especially those  closer  to the  Sun, contaminating  the
bright RC.  According  to the Besan\c{c}on model of  the Galaxy (Robin
et al. 2003),  the contamination from disk foreground  stars should be
small  for  both  RCs  (3$\%$  and  8$\%$  by  thin  and  thick  disks
respectively). The density of both disks  in the inner few kpcs of the
Galaxy, however,  has never  been constrained observationally,  so the
model predictions are extremely  uncertain on this point. Recently, the 
Besan\c{c}on model has been updated (Robin et al. 2012) including two 
populations in the inner bulge, and despite the differences with the old 
model (both for the bulge component but also the adjustments to the discs, 
most notably the thick disc), the figures for the contaminations are 
similarly low: 3\% and 7\%  contamination from the thin and thick disc, 
respectively,  for the BRC. While a contamination of 4\% and 7\% were 
estimated for the FRC. In order to estimate the contamination fractions 
quoted above we
split (arbitrarily) the single RC predicted in  two components,  each one
containing  half of the  bulge RC  stars, and  counted the  model disk
stars  in  two  boxes with  the  magnitude  and  color limits  of  our
spectroscopic targets.  Overall, we can expect that the stars at RV
close to  zero might  suffer from some  degree of  disk contamination,
hard to quantify here.

A second source of contamination comes from the bulge RGB stars in the
two overdensities  along the  line of  sight, because the
upper RGB  stars of  the far  arm (that corresponds  to the  faint RC)
overlap the bright  RC, whereas the lower RGB stars  of the nearby arm
contaminate the faint RC selection box. Furthermore, other bulge stars
not associated with the two arms (overdensities) might also exist along
the line of sight.

All  together, these contaminations  and cross--contaminations  act in
such a  way that each selection box  on the CMD will  not include purely
stars at the distances of the overdensities identified by the two RCs,
with the effect of smoothing  the features of the two RV distributions
seen in Figure~~\ref{rv},  and making them more similar  to each other
than they actually are.

An estimate of  the fraction of RGB stars included in  each of the two
RC boxes  has been calculated by  generating a synthetic  CMD from the
BASTi library of  stellar models, assuming an age of  10 Gyr and solar
metallicity.   Then  the  cross--contamination  has   been  evaluated
considering the  0.5 mag difference in distance  modulus ($\sim$2 kpc)
between the two overdensities and  counting the fraction of RGB and RC
stars  that fall  in each  selection  box. After  rescaling these  two
fractions by taking into account the cone effect, we finally estimated
a cross--contamination fraction of $\sim 18\%$ in both RCs.

It  is important  to note  that  the above  mentioned effects,  namely
foreground and cross--contamination,  are {\it certainly} present, even
if   their  impact  is   hard  to   precisely  quantify   here.   This
consideration supports  the assumption, that we make  in the following
analysis, according to  which the near arm is  dominated by stars with
mostly negative  radial velocities, roughly  centered at  $V_{\rm Helio}=-80$
km/s, whereas the  far arm is dominated by  stars with mostly positive radial
velocities, roughly centered at $V_{\rm Helio}=+80$ km/s.

\section{Proper Motions}

\begin{figure}
   \centering
   \includegraphics[width=9cm]{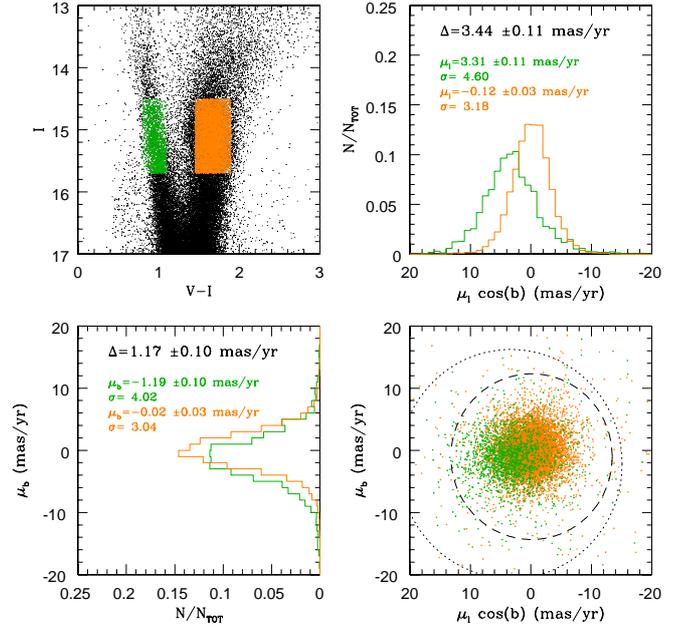}
      \caption{Proper motions measured at $(l, b) = (0, -6)$ field. RGB bulge stars and disk 
      dwarfs are shown with orange and green colors respectively. Dotted and 
      dashed circles in $\mu_{l}$ \textit{vs} $\mu_{b}$ plot show the $3 \sigma$ 
      data selection used for disk and bulge stars, respectively. The proper 
      motion distribution derived from this selection show a systematic
      difference between disk and bulge population, in agreement with 
      Clarkson et al (\cite{clarkson08}). 
      }\label{pmdisk}
   \end{figure}

\begin{figure}
   \centering
   \includegraphics[width=9cm]{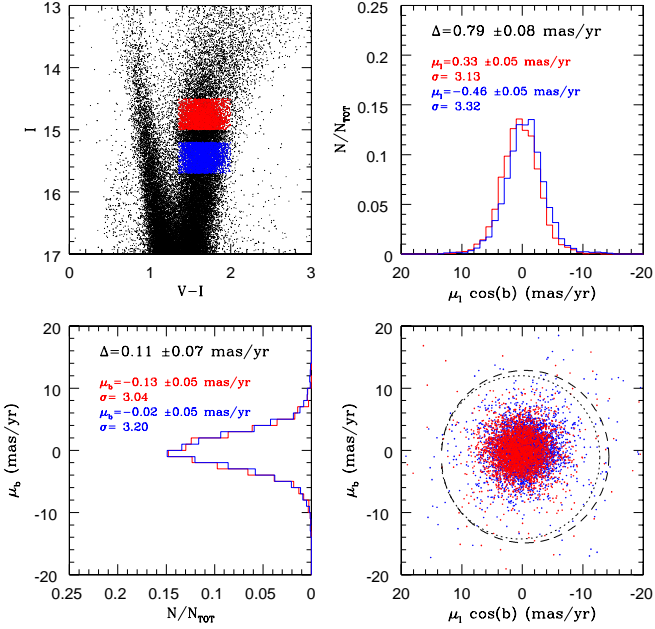}
      \caption{Proper motion measured at $(l, b) = (0, -6)$ field for 
      bulge RC stars. The 
      blue and red dots identify the selected bright RC (BRC) and faint RC (FRC) 
      stars on the CMD, respectively. A  $3\sigma$ selection 
      (lower right panel) in $\mu_{l}$
      and $\mu_{b}$ plane is used to define the sample further analysed with 
      histograms. Lower left panel shows $\mu_{b}$ distribution, while upper right 
      panel compares $\mu_{l}$ distributions for the bright (red) and faint 
      (blue) RC stars. 
      The mean values of $\mu_{l}$ and $\mu_{b}$ for the BRC and FRC samples 
      are given in the respective panels showing the distributions, and we 
      also provide the mean differences in
      proper motion for the two RC populations. 
      }\label{pmbulge}
   \end{figure}

\begin{figure}
   \centering
   \includegraphics[width=9cm]{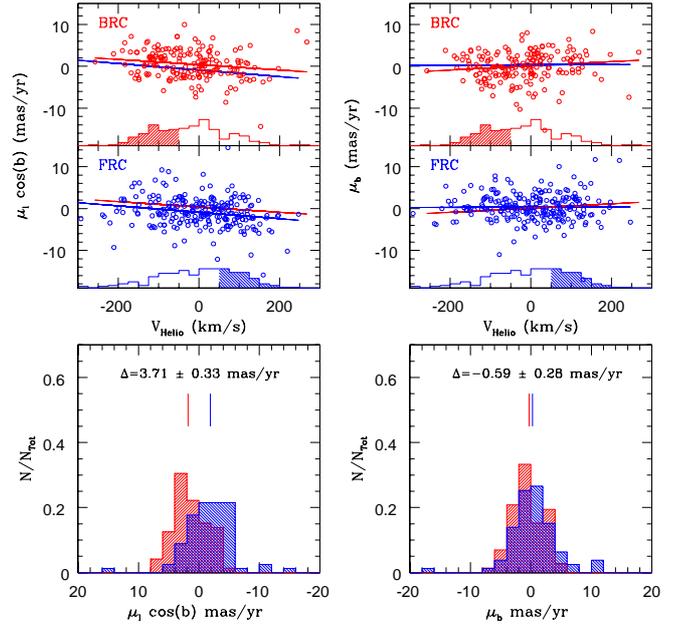}
      \caption{\textit{Upper panels}: PMs versus RVs for  BRC and FRC 
      sample. Colour lines (red for BRC and blue for FRC) shows the linear 
      regression for our data in the respective sample. As is shown in
      the right--upper panel, no significant difference in $\mu_{b}$ 
      versus RV is observed, while for $\mu_{l}$ a common trend with 
      RV is found 
      (left/upper panel). Under each scatter plot is also shown the
      scaled RV distribution for each sample, in order to define a 
      \textit{pure} BRC and FRC samples (red and blue dashed regions, 
      respectively). \textit{Lower panels:} Histograms for observed proper motions 
      and the difference in the median for 
      selected stars in each \textit{pure} sample defined above.     
      }\label{ppm_rc}
   \end{figure}

In  order to assess  the precision  of the  proper motions  derived as
described  in Section 2.2  we took  advantage of  the presence  in our
$(l,b)=(0,-6)$ field of the globular cluster NGC~6558. To this end, we
select the stars within a radius  of 2 arcmin from the cluster center 
(Half--light radius $\sim 2.15$ arcmin, taken from Harris \cite{harris96}). 
Due to the lower  metallicity of NGC~6558 ([Fe/H]$=-0.97\pm0.15$) 
with trespect  to the  bulge, it is  possible to select  a clean  sample of
clusters stars  from the observed CMD selecting stars on  the 
upper RGB ($I<16.5$) that are bluer than the bulge RGB. Cluster
stars are expected to have an internal velocity dispersion of $\sim$10
km/s (Pryor \& Meylan 1993), that, at the  distance of NGC~6558 
corresponds to a proper
motion  dispersion of  0.3  mas/yr,  too small  to  be measured  here.
Therefore,  we  attribute the  observed  dispersion  of cluster  stars
completely to observational errors.  After a 3 sigma-clipping, cluster
stars  show  an observed mean  proper  motion  of 
$(\mu_{\rm  l}\cos(b),\mu_{\rm b})=(0.30  \pm   0.14,  -0.43\pm  0.13)$,  
with   a  dispersion  of
$(\sigma_l \cos(b),\sigma_b)=(1.8,1.7)$ mas/yr.  The latter is,
therefore,  the observational  error  we will  use  to deconvolve  the
observed proper motion dispersion of bulge stars. Note that this is an 
upper limit to our real error, because cluster stars are more crowded 
than bulge stars, and because NGC6558 happens to fall at the edge of 
chip \#6.

Another test  to assess the accuracy  of our catalogue is  to separate a
pure disk  population, selected on the  blue main sequence  in the CMD
(as shown  in Figure~\ref{pmdisk}) and the  bulge population selected
along the  RGB near the RC.   The proper motion diagram  for these two
samples,  in galactic coordinates,  shows that  there is  a difference
between the mean PM of  the disk and bulge populations. After applying
a 3-sigma  clipping to both  distributions, the disk-bulge  offset was
found  to  be  $(\Delta\mu_{\rm l}\cos(b),\Delta\mu_{\rm  b})=(3.44\pm
0.11,  1.17\pm  0.10)$  mas/yr, which  compares favourably to  $(3.21\pm
0.15,0.81\pm 0.13)$  found by  Clarkson et al.   (2008) in a  field at
$(l,b)=(1.25,-2.65)$.  Additionally, if we deconvolve the bulge proper
motion dispersion from the  observational error as estimated above, we
obtain  $(\sigma_l  \cos(b),\sigma_b)=(2.62, 2.52)$  mas/yr,  in
excellent     agreement    with     the    values     of    $(\sigma_l
\cos(b),\sigma_b)=(2.64,2.40)$ mas/yr measured by Rattenbury et
al.    (2007)    in   two   fields    at   $(l,b)=(-0.25,-5.70)$   and
$(l,b)=(-0.14,-5.91)$.

The two epoch images  analyzed  here had not been obtained with the
purpose  of  deriving  proper  motions.  Therefore  the  observational
strategy (dithering, etc.)  was not optimized for this  kind of study.
Nevertheless,  the  measurements  reported  above confirmed  that  the
derived proper  motions, although  not very precise,  should be  good
enough to allow us to identify differences in the mean proper
motion of the  two bulge overdensities traced by  the bright and faint
RCs.

The   proper  motion   of  stars   in  the   two  RCs   is   shown  in
Figure~\ref{pmbulge}.  The bright RC contains stars with $14.5<I<15.0$
and  $1.35<V-I<2$, while  the faint  RC stars  have  $15.2<I<15.7$ and
$1.35<V-I<2$   as    illustrated   in   the    upper/left   panel   of
Figure~\ref{pmbulge}, where the two groups have approximately the same
number of stars. There is  a small but statistically significant shift
$\Delta\mu_{\rm l}{\rm cos}(b)=0.79\pm0.08$ mas/yr 
in  the  longitude  proper   motions  between  the  two  distributions
($\sim$100\%  confidence from  a KS  test), whereas  the  two latitude
proper motion distributions are instead barely distinguishable.

For  the   spectroscopic  targets  it  is  possible   to  combine  the
information from  the PMs with that  from RVs, thus  deriving 3D space
velocity for individual stars.  The upper panels of Figure \ref{ppm_rc}
show  that both  bright and  faint RC  stars share  a common  trend in
$\mu_{\rm l}$ \textit{vs} $\rm V_{\rm Helio}$  which is  not observed 
in  $\mu_{\rm b}$ (Pearson correlation coefficient of $\sim-0.2$ and $~\sim 0.05$, respectively).  To emphasize
this behaviour a  linear regression between PMs and  RV was calculated
for each sample and over--plotted on the data points.

Following  the arguments given  in Section~3,  the cleanest  sample of
bright RC stars is the  one with negative radial velocities, while the
cleanest sample of  faint RC stars has positive  radial velocities. If
these arguments are correct, then the stars we want to probe should be
at  the extreme of  the two  radial velocity  distributions. Arbitrary
selections  of  stars  with   $V_{\rm  Helio}<-50$  km/s  and  $V_{\rm
  Helio}>+50$   km/s  were  made   to  the   bright  and   faint  RCs,
respectively, and they  are shown at the bottom  of each scatter plot,
where small versions  of the RV distributions are  also shown. The PMs
of  these stars  are compared  in  the two  lower panels  in the  same
figure, showing a significant offset  in $\mu_{\rm l}$, while the mean
$\mu_{\rm b}$ values are virtually identical.
 
These  results can be  interpreted as  follows.  In  the near  and far
overdensities, we  observe stars on  radial orbits ($V_{\rm  Helio}$ 
non-zero).  An
excess  of stars  approaching  the Sun  is present  in the  near
overdensity, while an excess of stars receding from the Sun is present
in the far  overdensity. By itself, this is an  indication of stars in
streaming motions along  the {\it bar}.  In our bulge  the bar has been
proved  to be  X-shaped.  Structures of this kind have been proposed to 
form by buckling  and bending of the stellar distribution  were stars 
are re-arranged on banana--orbits, i.e., the family of x1v1 orbits 
according to Patsis, Skokos \& Athanassoula (2002), that  look either 
like a `frown'  ($\frown$) or a
`smile' ($\smile$)  when viewed edge on.   This is the  main family of
stable, periodic orbits forming a peanut-shaped, or an X-shaped bulge.

Stars near the velocity inversion points of such orbits, which are the
ones  that we  are  probing at  $b=-6$,  are expected  to go  in 
opposite radial, as well as opposite longitude directions,
if, as  in our case, the line  of sight crosses two  opposite sides of
the  orbit\footnote{The line of  sight at  ($l,b$)=($0,-6$) intercepts
  the  two overdensities  formed  by the  family  of {\it  frown}-like
  orbits. In particular, it touches the  West side of the near part of
  the  {\it  frown}, and  the  East side of  the  far  one.}.  On  the
contrary, since  both overdensities, at  this latitude, lie  below the
plane, they are expected to share a common mean $\mu_b$, as observed.

Our  result    differs   from   that   of   De    Propris   et   al.
(\cite{depropris11}),  who   found  no  differences   between  the  RV
distribution of bright  and faint RC stars in a fields at $b=-8$. Although we do not have a conclusive explanation
for this discrepancy,  a few factors can account for  at least part of
it.   De  Propris  et  al. (\cite{depropris11})  have  selected  their
targets from the 2MASS CMD.  It is clear from their Figure~1 that they
have larger contamination  from disk main sequence stars  in the faint
RC, and possibly  also from the bright RC (due to  the larger error on
individual  magnitudes of  2MASS versus  WFI photometry).   A slightly
larger error on the magnitudes of individual stars in 2MASS versus WFI
photometry, and a slightly larger error on the radial velocities (from
their lower S/N and lower resolution spectra) might also contribute to
mask out the features we found in the faint RC distribution.

Radial velocities for stars in the  two RCs visible on the bulge minor
axis were recently derived also  by Uttenthaler et al. (2012) and Ness
et al. (2012). The latter compares the measured distributions with the
prediction of a bulge model  by Athanassoula (2003). The model was not
optimized  to reproduce  the  Galactic bulge,  hence  cannot be  taken
quantitatively, but  it is a model  of a peanut shaped  bulge that, at
least qualitatively,  reproduces the two  observed overdensities along
the lines of sights on the minor axis. The model predicts the presence
of  a  clear  asymmetry,   in  the  radial  velocity  distribution  at
$b=-5$, similar  to the  one observed by  Ness et al.  (2012) at
this latitude,  and to  the one presented  here. At  larger latitudes,
however, the asymmetry  becomes much weaker in the  model, and also in
the  observations  by  Ness   et  al.  (2012)  at  $b=-7.5$  and
$b=-10$. The fact that the asymmetry is predicted to weaken at 
high latitudes might, perhaps, explain the similarity between the RV
distributions  of bright  and  faint  RCs observed  by  de Propris  et
al.  (2010)  and  Uttenthaler  et   al.  (2012).

We can conclude that we  have identified stars, in  the near and
far overdensities  of the lower part  of the X-shaped  bulge, whose 3D
kinematics  are  qualitatively  consistent  with   the  prediction  of
dynamical models producing a peanut-shaped, or an X-shaped, bulge from
the buckling instability  of a bar. These are not  {\it all} the stars
we see  in the near and far  overdensities traced by the  near and far
RCs.  In each  of the two RCs there are indeed  also stars sharing the
same 3D  velocities, as well  as a small  number of stars  with radial
velocities  opposite to the  main stream.  These can  be qualitatively
interpreted as  stars in more stochastic  (spheroid-like) orbits, plus
some degree of cross contamination (cf Section~3).

\section{Space Velocities}

\begin{figure}
   \centering
   \includegraphics[width=9cm]{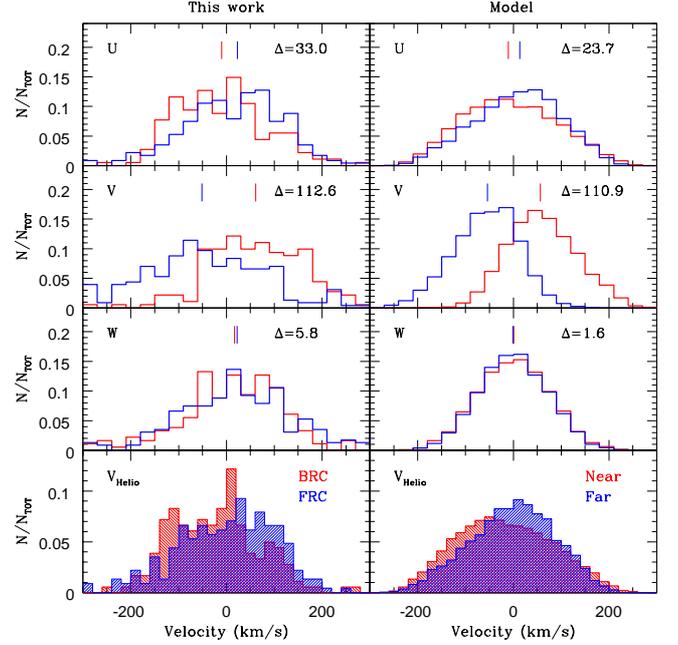}
      \caption{BRC and FRC are compared with respect the kinematical model 
      for a strong Boxy--Peanut bulge (Debattista et al. \cite{Debattista+05}). 
      From the model, two samples were selected  around the two 
      overdensities formed by the Near (red) and Far (blue) 
      arms of the Boxy--Peanut stellar distribution, in the line of 
      sight for ($l,b$)=($0,-6$). Colour lines  over 
      U, V and W histograms correspond to the median value for each 
      distribution.
      }\label{model_comp}
   \end{figure}

Space velocities  were obtained for all the  spectroscopic targets, in
the $U,V,W$  Galactic Cartesian system following Johnson \& Soderblom (1987),
assuming that  stars in the  bright and faint  RC are on average  at a
distance of $6.5$ and $8.5$  kpc from the Sun, respectively. Because our proper 
motions are relative to mean position of the bulge stars instead of quasars 
or distant galaxies, it is necessary to correct them by the relative motion of 
the galactic center with respect to an inertial frame before computing 
$U$, $V$ and $W$ velocities. Sumi et al. (2004) determine 
this correction as:\footnote{$\mu_{\alpha *} = \mu_{\alpha}\cos\delta$} 
\begin{equation}
{\mu_{\alpha *}}_{\rm INERT} = {\mu_{\alpha *}}_{\rm OBS} - {\mu_{\alpha *}}_{\rm GC} +  {\mu_{\alpha *}}_{\rm GC,INERT}
\end{equation}
\begin{equation}
{\mu_{\delta}}_{\rm INERT} = {\mu_{\delta}}_{\rm OBS} - {\mu_{\delta}}_{\rm GC} +  {\mu_{\delta}}_{\rm GC,INERT}\;,
\end{equation}
where the subscript $\rm OBS$ and $\rm GC$ refer to observed PM of individual 
stars and the mean PM of the galactic bulge respectively. The quantities 
(${\mu_{\alpha *}}_{\rm GC,INERT},{\mu_{\delta}}_{\rm GC,INERT})=(-2.93,-5.17)$ 
mas/yr correspond to the expected motion of the galactic center with respect 
to the inertial frame. The spatial velocities were corrected from the peculiar 
solar motion $(u,v,w)_{\odot}=(11.1,12.24,7.25)$ km/s 
(Sch\"onrich, Binney and Dehnen 2010), considering the solar 
circle radius and Local Standard of Rest as $R_{\rm GC}=8.0$ kpc and 
$\rm LSR_{\odot}=220$ km/s respectively. We adopted  the usual notation
where  $U$  is the  radial  component  along  the star-Galactic  center
direction, positive  towards the Galactic center; $V$  is the peculiar
rotational velocity of each star with respect to its Local Standard of
Rest,  positive in  the sense  of the  Galactic rotation;  $W$  is the
vertical velocity component, positive towards the North Galactic pole.


\subsection{Model Comparison}

\begin{table*}
\caption{Statistics for velocities presented in Fig. \ref{model_comp}, 
in km/s.}
\label{t1}
\centering
\begin{tabular}{l l l l l l l l l l l}
\hline\hline
& & $V_{\rm Helio}$ & U & V & W & & $V_{\rm Helio}$ & U & V & W\\
\hline
\\
& & BRC $(N=181)$: & & & & & FRC $(N=227)$: & & &\\
\hline
Mean     & & $-22.3 \pm 6.9$ & $-10.0 \pm 7.0$ & $65.9 \pm 8.2$ & $13.8 \pm 8.4$ &     & $1.0 \pm 6.7 $ & $15.0 \pm 6.8$ & $-42.6 \pm 10.4$ & $26.6 \pm 8.9$ \\     
$\sigma$ & & $62.1 \pm 4.9 $ & $64.5 \pm 4.9 $ & $68.5 \pm 5.8$ & $66.4 \pm 5.9$ &     & $72.3 \pm 4.8$ & $70.4 \pm 4.8$ & $86.3 \pm 7.3  $ & $77.1 \pm 6.3$ \\    
Median   & & $-21.4        $ & $-9.9         $ & $61.4        $ & $16.8        $ &     & $9.5         $ & $23.1        $ & $-51.2         $ & $22.6        $ \\
\hline  
\\
& & Near $(N=9590)$: & & & & & Far $(N=7349)$: & & &\\
\hline
Mean     & & $-17.0 \pm 1.0$ & $-5.9 \pm 1.0 $ & $60.4 \pm 0.7$ & $0.6 \pm 0.8 $ &     & $-2.7 \pm 1.0$ & $8.1 \pm 1.1 $ & $-57.7 \pm 0.8$ & $-0.2 \pm 0.8 $\\
$\sigma$ & & $74.0 \pm 0.7 $ & $72.0 \pm 0.7 $ & $49.4 \pm 0.5$ & $51.9 \pm 0.5$ &     & $61.3 \pm 0.7$ & $63.8 \pm 0.7$ & $46.8 \pm 0.6 $ & $49.2 \pm 0.6 $\\
Median   & & $-21.7        $ & $-10.3        $ & $56.4        $ & $0.9         $ &     & $1.5         $ & $13.4        $ & $-54.5        $ & $-0.7         $\\
\hline
\end{tabular}
\end{table*}

We compare our observations to the high resolution model R1 of
Debattista et al. (\cite{Debattista+05}).  Here we briefly describe 
this model and the
method we used for comparing with the observational data.  This model
starts out with an exponential disk with a ratio of
scale-height to scale-length, $z_d/R_d = 0.05$.  The disk has a
Toomre-$Q = 1.2$ and is immersed in an isothermal halo with
scale-length $r_h = 3.3 R_d$.  The model has $7.5 \times 10^6$
particles; it was evolved on a cylindrical grid code described in
Sellwood \& Valluri (\cite{Sellwood_Valluri97}).  The model 
forms a bar and experiences a
violent buckling instability leading to the formation of a B/P shape.
The formation of the B/P shape is presented as an animation in
Debattista et al. (\cite{Debattista+2006}) 
(where it is referred to as model L2).

In order to compare the model with the Milky Way, we need to rescale
from natural units in which $R_d = M_d = G = 1$ (where $M_d$ is the
disc mass) to physical units and rotate the model to reproduce the 
observed inclination bar in the Galaxy. To scale sizes, we use units 
of the bar radius, placing the observer at two bar radii from the centre 
of the galaxy, with the bar extending to $\sim 1.45 R_d$ of the 
initial disc.  We scale velocities by multiplying the unit velocity 
by 250 km s$^{-1}$. The bar in the model was rotated to an angle 
of 20 degrees relative to the Sun-Galactic centre direction. 
We selected particles from both sides of 
the center of the model, at an observing window of 
($l,b$)=($0,-6$). The selection of particles was 
done based on their spatial distribution 
along the $x$ coordinate. Two overdensities were observed, corresponding with 
the two arms of the ``X'' (B/P--shape). Around each peak, Near and Far samples
were defined  in order to better represent the 
position of the clumps, avoiding background and foreground 
contamination. Figure \ref{model_comp} shows a qualitative comparison 
between the observations and the model, in which the median values for 
the BRC/Near and FRC/Far samples and in particular their differences 
are consistent. Table \ref{t1} shows the mean, dispersion 
and median for each velocity component present in Fig. \ref{model_comp}. 
As is possible to observe, most of the velocities predicted 
by the model are consistent with the observed ones (within the errors) 
except for W, which is hotter than the model in $\sim 20$ km/s. A 
mismatch in the specific value of the mean and dispersion for the velocity 
distributions (instead the difference) are not unexpected, due the model 
has not been built particularly for the Milky Way. Additional 
analysis will be presented in Gardner et al. (in prep).

It is worth emphasizing that this consistency of our data with the model 
does not necessarily prove that the model fully represent the formation 
history of the MW bulge. Indeed, in this class of purely N-body models 
(e.g., Debattista et al. 2006; Shen et al. 2010) the gas contribution 
of the overall event of bulge formation is not taken into account. Instead, 
when the bulk of the bulge stars formed, some $\sim 10$ Gyr ago, the Galaxy 
was likely to be very gas rich, with a gas fraction of $\sim 50\%$, as 
indicated by direct observations of star forming galaxies at the corresponding 
lookback time (i.e., at $z\simeq 2$, see Tacconi et al. 2010; Daddi, et al. 2010).

\section{Metallicity distribution}

Calcium  II Triplet  (CaT) metallicities  for both  the IMACS  and the
GIRAFFE spectra have been derived  using the calibration by V\'{a}squez et
al.  (in prep).  The  latter has been constructed by observing a
sample  of  RC and  RGB  stars in  Baade's  Window,  with the  GIRAFFE
spectrograph, in the LR08 setup,  centered at 8817 $\AA$. GIRAFFE high
resolution spectra had been previously  obtained for the same stars by
our group,  and [Fe/H] measurements  based on such spectra 
are  presented in  Hill et  al.  (2011).   The comparison  between the
equivalent  widths (EWs) of  CaT lines  in the  LR08 spectra  and the
[Fe/H] abundances, obtained from the EWs of individual FeI lines, yields
a linear relationship that extends  up to [Fe/H]$\sim +0.5$~dex, and it
is fully compatible  with the analogous relations obtained  by Cole et
al.   (2004)  and Warren  et  al.  (2009).   The  details  of the  new
calibration will be discussed in  a separate paper.  We emphasize here
that  it   has  been  constructed  precisely  for   bulge  giants  and
particularly tested for RC stars.

In  order  to  convert  EWs  into metallicities  it  is  necessary  to
calculate  the so  called {\it  reduced} EWs,  a  parametrization that
removes the effects  of temperature and gravity on the EWs, defined as
\begin{equation}
W'=\Sigma EW + \beta  (K-K_{RC}),
\end{equation} 
where $K_{\rm RC}$ is the mean $K_s$ magnitude of the RC. In our case,
$K_{\rm RC}$  was obtained independently  for the bright and  faint RC
using  the photometric  catalogue from  the VVV survey  
(Saito et  al. 2012)  and
extinction maps  from Gonzalez et  al.  (2012).  The  quantity $\Sigma
EW$ is the sum of the  EWs of the three calcium lines. The measurement
of the  EWs of  CaT lines is  rather difficult  for stars of  solar and
super-solar metallicity,  due to the presence of  many small molecular
lines  that contaminate  the {\it  pseudo}-continuum. In  the  case of
IMACS spectra  the task  is further complicated  by the fact  that the
dispersion axis goes across 4 different chips of the mosaic, with gaps
between them.   For this reason,  as well as due to  lower S/N of  the IMACS
spectra  with  respect  to  the   GIRAFFE  ones  used  to  derive  the
calibration, the  metallicity derived from IMACS spectra  are not very
accurate. In particular, for four stars in common between the IMACS
and GIRAFFE samples, the derived  metallicities show both an offset 
(of $\sim$0.2 dex) with a mild trend with metallicity.
Rather than relying on just four stars to measure and correct for this
systematics, we prefer to analyse here only the metallicities for the
IMACS targets, without combining them with the GIRAFFE ones.

One might argue that we are ignoring the highest quality data (GIRAFFE
sample)  keeping only  the  lower quality  ones  (IMACS sample).  This
choice  is due  to  the larger  number  of IMACS  targets,  and it  is
justified by the assumption that  the impact of the systematics in the
measurements  of EWs  in IMACS  would be  small, if  metallicities are
considered only in a {\it relative} scale. The IMACS metallicities will
be labelled as {\it instrumental}, hereafter, in order to emphasize
that we have indications that they might not be accurate, in an absolute 
scale. Moreover, these CaT metallicities are used uniquely to check 
whether the bright and faint clumps 
have consistent or (slightly) different metallicity distributions.

\begin{figure}
   \centering
   \includegraphics[width=9cm]{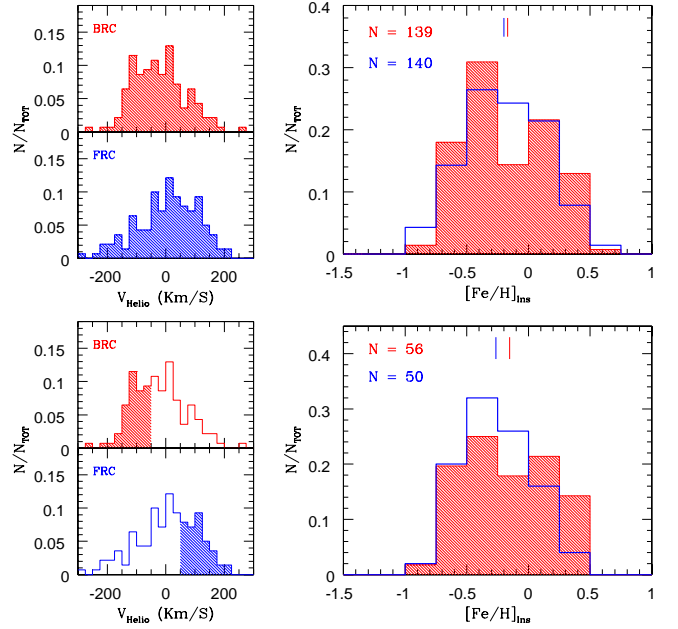}
      \caption{Instrumental metallicity distribution for IMACS data. Selecting 
      all the BRC and FRC in whole RV regime (upper--left panels), the 
      [Fe/H]$_{\rm Ins}$ distribution plotted in upper--right panel do not show 
      a statistical difference. Nevertheless when a selection is done in the 
      same way as Figure \ref{ppm_rc} (lower--left panels), a shift in the 
      [Fe/H]$_{\rm Ins}$ distribution is observed in lower--right panel. 
      The star number is also indicated in each case following the key color 
      adopted in this paper and the corresponding selection showed in left panels. 
      }
      \label{met}
   \end{figure}

Figure~\ref{met} shows the metallicity  histogram for the target stars
in the  bright (red) and the faint  RC (blue). As already  found by de
Propris et al.  (\cite{depropris11}), the  two samples do not show any
difference  in their  mean  metallicity. However,  if  we select  only
bright  RC stars  with $V_{\rm  R}<-50$ km/s  and faint  RC  stars with
$V_{\rm R}>+50$ km/s, as already done for the PMs, then a difference of
$\sim0.11$~dex is found between the mean [Fe/H] of the two samples.  A
small difference between  the near and the far  overdensities of bulge
stars  at  ($l,b$)=($0,-6$)  is  consistent  with the  presence  of  a
gradient  along  the  minor  axis,  as  observed  by  Zoccali  et  al.
(\cite{zoccali08}), Johnson et al. (2011), and Gonzalez et al. (2011).
Indeed, the line  of sight crosses the near  and the far overdensities
at different distances  from the Galactic center. In  order to estimate
the  expected   difference  in  metallicity,  the   distance  to  each
overdensity   was   determined    following   McWilliam   \&   Zoccali
(\cite{mcwilliam10}),  finding  a  $\Delta  Z  \sim$200  pc  in  their
distance to the Galactic  plane.  Assuming a $0.6$~dex/kpc gradient, a
difference  of $0.12$~dex  in  metallicity would  be expected  between
stars in the  two RCs.  Clearly, this comparison  is very preliminary,
because the radial velocity cut is arbitrary, and because the gradient
quoted  in the literature  is based  on samples  of stars  without line of sight 
distance  information, i.e.,  summing  up -in  an unknown  proportion-
stars  in  the near  and  far  overdensities.   In summary,  there  is
evidence, in the literature, of the presence of a metallicity gradient
along the  bulge minor axis. While the  value of this  gradient is
quite uncertain, here we found  that adopting a radial velocity cut
that  minimizes cross contamination  a metallicity  gradient appears
that is consistent with that reported by Zoccali et al. (2008).

\begin{figure}
   \centering
   \includegraphics[width=9cm]{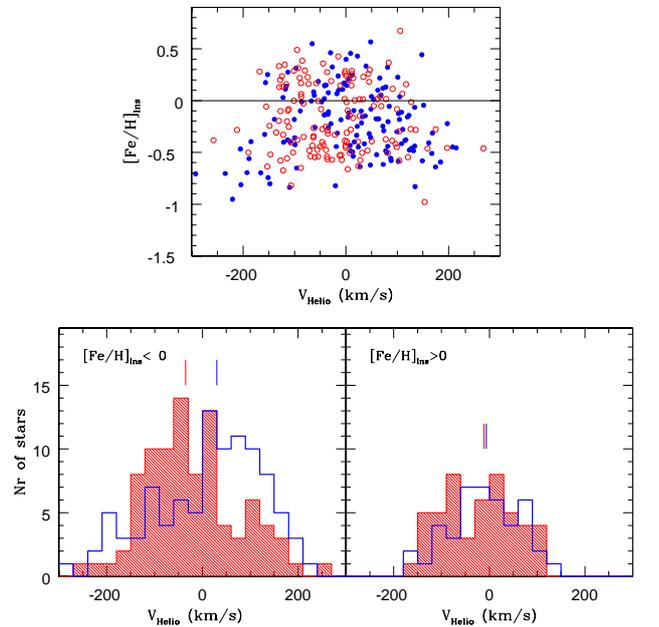}
      \caption{[Fe/H] versus RV for BRC and FRC samples observed with 
      IMACS. The histograms show an interesting difference when the sample 
      is divided in metal poor and metal rich.  
      }
      \label{met_vel}
   \end{figure}

Finally,   another  interesting   piece  of   evidence  is   shown  in
Figure~\ref{met_vel}.   The metallicity  of the  bright (red)  and faint
(blue) RC stars  is shown here as  a function of RV. A  glance at this
plot shows some segregation of the faint RC stars on the right half of
the plane,  while the  bright RC ones  are preferentially on  the left
side.  The  additional evidence appearing here is that
this segregation is stronger  for metal poor stars ([Fe/H]$<0$), shown
in  the  lower  left  panel,  than  it is  for  the  metal  rich  ones
([Fe/H]$>0$), shown in the lower right panel. Surprisingly, this would
argue in  favor of  the metal poor  stars being preferentially  in the
more elongated orbits, defining the X-shape, while the most metal rich
ones are preferentially found on axysimmetric orbits, centered at zero
RV and with lower velocity dispersion.

Ness et  al. (2012) also found  that metal rich RC  stars have smaller
velocity dispersion than metal poor  ones, consistent with what we see
in Fig.~8.   They also found that the  magnitude (distance) separation
between  the two  RCs is  stronger for  metal rich  stars.  The latter
evidence cannot  be checked here because our  initial target selection
excluded stars  in between  the two RCs.  Our results that  the radial
velocity asymmetry is stronger for  the metal poor half of the sample,
however,  suggests  that  the  magnitude splitting  should  be  there,
perhaps even stronger, for these stars.

\section{Conclusions}


We have analysed a sample of 454 bulge stars equally distributed between the
bright and the faint RCs of a bulge field at ($l,b$)=($0,-6$). The two
RCs are used here as tracers  of the near and far overdensities of the
X-shaped bulge, crossed by the  line of sight at these coordinates.  We
obtained radial velocities  and proper motions for all  the stars, and
CaT metallicities for a subsample of 352 stars.
 
We also measured the proper motion  of NGC~6558, finding an observed 
mean proper motion of 
$(\mu_{\rm  l}\cos(b),\mu_{\rm b})=(0.30 \pm 0.14, -0.43\pm
0.13)$, with  a dispersion of  $(\sigma_l \cos(b),\sigma_b)=(1.8,1.7)$
mas/yr.   This is the first PM measurement for NGC~6558.

An excess  of bulge stars  in elongated orbits  ($|V_{\rm R}|>50$  km/s) has
been  found  in  both   RCs,  with  the  near  overdensity  containing
preferentially stars  with negative RVs, while the  far one contains
preferentially stars with positive RVs. The 3D velocities of these
stars are  qualitatively  consistent with the predictions of $N$-body models by 
Debattista et al. (2005) with a strong X-shaped bulge.

Interestingly,  not all  the stars  in the  two RCs  are  in elongated
orbits. Roughly half of the stars  in both the bright and the faint RC
share the  same 3D velocity,  with mean RVs  and PMs centered  at zero.
They  seem  to  be  stars  in axysimmetric  orbits  that  are  somehow
coexisting with  those in elongated orbits. Whether  the two kinematic
groups  share the  same  origin, i.e.,  they  both belong  to the  same
Galactic  component  is  impossible  to  say with  the  present  data.
Nevertheless, it  is interesting to  note that current  models predict
that  the presence of  a bar  should clean  up the  inner region  of a
Galaxy, i.e.,  no stars in  axysimmetric orbits should be  left there,
regardless of their origin.

Metallicity measurements provide further some evidence for the stars
in elongated orbits, i.e., those with the largest difference in RV
and PMs, being preferentially the ones with sub-solar metallicities, 
while the more metal rich ones in both the near and far overdensity
share the same 3D motions, with RV and PMs centered at zero, and
a smaller radial velocity dispersion. This finding, if confirmed,
would be at odds with the suggestion by Babusiaux et al. (2010) that
metal poor bulge stars have spheroid-like kinematics, while the
metal rich ones have a significant vertex deviation, typical of 
stars in bar-like orbits.

Clearly,  all the  evidence presented  here is  based on  a relatively
small number of stars, and  needs to be confirmed with larger samples,
such as  the ESO--Gaia Public Spectroscopic Survey (PIs: Gilmore  \& Randich),
the ARGOS  survey (PI: Freeman),  ESO Large Programme  187.B-0909 (PI:
Zoccali) and  the APOGEE survey  (PI: Majewski).  Indeed,  the present
pilot  study  demonstrates  the  amount  of information  that  can  be
acquired by combining spectra with multiepoch photometry for red clump
stars, for which distances are known.  For example, with the ESO Large
Programme  we expect to  soon be  able to  use this  kind of  data for
significantly larger samples of stars along many different bulge lines
of sight. As  a result, important constraints will be  set to state of
the art  dynamical models, with  the ultimate aim of  establishing the
relative roles of the processes that contributed to make the Milky Way
bulge as we see it today.


\begin{acknowledgements}
  We thank Santino Cassisi for calculating the expected fraction of red 
  giant branch contamination in the red clump, from his models.
  SV and MZ acknowledge  support by Proyecto Fondecyt Regular 1110393.
  While working  at this  paper, MZ enjoyed  a sabbatical year  at the
  INAF Osservatorio Astronomico di  Bologna, and the European Southern
  Observatory in  Garching. Both  institutions are warmly  thanked for
  their kind hospitality.  A fellowship from the John Simon Guggenheim
  Memorial  Foundation has  partly financed  this stay.  This  work is
  supported  by  the  BASAL  Center for  Astrophysics  and  Associated
  Technologies  PFB-06, the FONDAP  Center for  Astrophysics 15010003,
  Proyecto Anillo ACT-86 and by  the Chilean Ministry for the Economy,
  Development,   and  Tourism's  Programa   Iniciativa  Cient\'{i}fica
  Milenio through grant P07-021-F, awarded to The Milky Way Millennium
  Nucleus.
  
  MM was suppported by the IAC (grants 310394, 301204), the Education and
 Science Ministry of Spain (grants AYA2010-16717).
\end{acknowledgements}


\end{document}